\pdfoutput=1

\documentclass[twocolumn,showpacs,amssymb,aps,nofootinbib,floatfix,superscriptaddress]{revtex4-1}

\usepackage{epsfig}
\usepackage{float}
\usepackage{amssymb}
\usepackage{hyperref}

\newcommand{\be}{\begin{equation}}
\newcommand{\ee}{\end{equation}}

\newcommand{\bea}{\begin{eqnarray}}
\newcommand{\eea}{\end{eqnarray}}
\newcommand{\ba}{\begin{array}}
	\newcommand{\ea}{\end{array}}

\begin{document}

\title{Higher order cumulants of transverse momentum and harmonic flow in relativistic heavy ion collisions}

\author{Piotr Bożek}
\email{piotr.bozek@fis.agh.edu.pl}
\affiliation{AGH University of Science and Technology, Faculty of Physics and
Applied Computer Science, aleja Mickiewicza 30, 30-059 Cracow, Poland}

\author{Rupam Samanta}
\email{rsamanta@agh.edu.pl}
\affiliation{AGH University of Science and Technology, Faculty of Physics and
Applied Computer Science, aleja Mickiewicza 30, 30-059 Cracow, Poland}

\begin{abstract}
Higher order symmetric cumulants of  global collective
observables in heavy ion collisions are studied. The symmetric cumulants can be straightforwardly  constructed for scalar observables~: the average transverse momentum, the multiplicity, and the squares of   harmonic flow vectors. Third and fourth order cumulants are calculated in the hydrodynamic model. 
  A linear predictor of the average transverse momentum and harmonic
  flow coefficients in a collision is used 
to predict the values of the cumulants from the moments of 
the initial distribution.  The symmetric cumulants divided   by the averages (or the standard deviations) of the considered observables can be used as a fine
tool
to study correlations present in the initial state of the collision.
\end{abstract}


\maketitle

\section{Introduction}

The dense matter created in relativistic heavy ion collisions
expands rapidly.
The study of  the collective flow of the expanding matter,
modeled by the viscous hydrodynamics, can be used to study its properties
\cite{Ollitrault:2010tn,Heinz:2013th,Gale:2013da}.
One of the  difficulties  in this tasks lies in the uncertainty
on  the initial conditions for the hydrodynamic evolution. To reduce
the uncertainty in the prediction of the harmonic flow coefficients,
a number of
additional observables based on correlators involving four or more particles 
have been proposed to constrain the initial conditions.

The expansion in the directions
transverse to the beam direction  generates the collective transverse flow.
Of particular importance are the harmonic coefficients of the azimuthal
particle distribution reflecting the  azimuthal
asymmetry of the initial conditions.  The average transverse momentum of
particles emitted in a collision is a measure of the overall radial expansion.
The final transverse momentum is related to the size of the initial source 
\cite{Broniowski:2009fm}. Events with a smaller size of the interaction region
have larger energy density
gradients and  a larger transverse push is formed in the expansion. The scaled
transverse momentum fluctuations reflect, in the first approximation, the fluctuations of the initial size
and in the entropy deposition.

A finer tool to look at the correlations  of the collective flow observables in heavy ion collisions is given by the correlation
coefficient $\rho(p_T,v_{n}^2)$ between the average transverse 
momentum $p_T$  and the harmonic flow coefficient $v_n^2$ in an event
\cite{Bozek:2016yoj}.
The  correlation of the elliptic or
triangular flow with the average transverse momentum  is well described by the
 correlations present in the initial state of the hydrodynamic evolution
 \cite{Bozek:2020drh,Schenke:2020uqq,Giacalone:2020dln}. Thus, 
the correlation coefficient of transverse momentum $p_T$ and the elliptic or triangular 
flow coefficient can serve as a tool to study correlations present in the initial state. 
The correlation coefficient  has been measured for
Pb+Pb and p+Pb collisions by the ATLAS Collaboration \cite{Aad:2019fgl}.
The experimental data on $\rho([p_T],v_{2}^2)$  in  Pb+Pb collisions are qualitatively well described
by hydrodynamic models. On the other hand, the magnitude and the sign of the   correlation coefficient
$\rho([p_T],v_{3}^2)$ are not well  described in most of the calculations.

The study of the correlation coefficient between transverse momentum and elliptic flow
is particularly interesting in deformed nuclei. In
central collisions of deformed nuclei  the orientation of the colliding deformed nuclei leads to specific correlations between the initial size and the elliptic deformation  \cite{Giacalone:2019pca,Giacalone:2020awm}. Experimental results show a sensitivity of the correlation coefficient $\rho([p_T],v_{2}^2)$ to the nuclear deformation  \cite{ATLAS:2021kty,JiaIS2021}.
The  transverse   momentum-harmonic flow  correlation coefficient 
in peripheral events  can be sensitive  to initial momentum correlation
that would survive the hydrodynamic evolution \cite{Giacalone:2020byk}.

In this paper we explore higher order correlators of the harmonic
flow and of the average transverse momentum.  We study higher order symmetric cumulants \cite{Bilandzic:2013kga,Mordasini:2019hut,Moravcova:2020wnf}
of scalar quantities: the transverse momentum, multiplicity, and the squares of harmonic flow vectors of different order.
We present predictions for the third and four order
normalized symmetric cumulants for collisions of spherical Pb+Pb
and deformed U+U nuclei in the hydrodynamic model.  We propose  two different ways to normalize the higher cumulants, by the average or by the standard deviations of the observables taken in the cumulant. The calculations
are compared to the results obtained using a linear predictor based on   moments
of the initial density. The higher order cumulants calculated in the
hydrodynamic model could be
measured experimentally.

\section{Correlations of global observables in an event}

\subsection{Model}

Expanding the azimuthal dependence of particle distributions in Fourier series we can write 
\begin{equation}
\frac{dN}{dp d\phi}=\frac{dN}{2\pi dp}\left( 1+  2 \sum_{n=1}^{\infty} V_n(p) e^{i n\phi}\right) \ ,
\end{equation}
using the complex plane  notation for the transverse plane ($p$,$\phi$). 
In this paper we use boost invariant calculations and the results
should be compared to experimental results obtained  at  central rapidities in  heavy-ion collisions.
Both the transverse momentum distribution $\frac{dN}{dp}$ and the harmonic flow vectors $V_n$ fluctuate event by event.
In this paper we study event by event correlations between moments of the distribution $\frac{dN}{d p d\phi}$ integrated in a range of transverse momenta, the charged particle multiplicity
\begin{equation}
N=\int_{p_{min}}^{p_{max}} dp   \frac{dN}{dp}
\end{equation}
the average transverse momentum in an event
\begin{equation}
p_T=\frac {1}{N} \int_{p_{min}}^{p_{max}} p dp  \frac{dN}{dp}
\end{equation}
and the harmonic flow coefficients
\begin{equation}
V_n =  \frac{1}{N}  \int_{p_{min}}^{p_{max}}  dp  V_n(p) \frac{dN}{dp} . 
\end{equation}
The complex numbers $V_n=v_n e^{i n \Psi_n}$
describe the flow magnitude $v_n$  and the flow angle $\Psi_n$ for the harmonic flow of order $n$.
The average over a  sample of events in a given centrality bin is denoted by $\langle \dots \rangle$, e.g. the averages  of the multiplicity  $\langle N \rangle$ ,  transverse momentum  $\langle p_T \rangle$, and   flow harmonics 
\begin{equation}
v_n\{2\}= \sqrt{ \langle V_n V_n^\star \rangle}
\end{equation}
can be calculated.

We use the 2-dimensional  version of the  hydrodynamic code  MUSIC \cite{Schenke:2010nt,Schenke:2010rr,Paquet:2015lta}
with shear viscosity $\eta/s=0.08$ for the hydrodynamic expansion phase. For the initial conditions in Pb+Pb collisions at $\sqrt{s_{NN}}=5.02$~TeV
we use a two-component Glauber Monte Carlo model
\cite{Bozek:2019wyr}
to calculate the initial entropy density in each event. The details of the model for the initial density can be found
in \cite{Bozek:2016yoj}.
 The initial density  in the transverse plane $s(x,y)$ in each event can be characterized by specific moments,
 the eccentricities
 \begin{equation}
 \epsilon_ne^{i\Phi_n} = - \frac{\int rdr d\phi \  r^{n} s(r,\phi )e^{i n \phi  }}{\int rdr d\phi  \ r^{n} s(r,\phi )} \ ,
 \end{equation}
 the RMS radius
 \begin{equation}
 R^2=  \frac{\int rdr d\phi \  r^{2} s(r,\phi )}{\int rdr d\phi  \  s(r,\phi )} \ ,
 \label{eq:rms}
 \end{equation}
and the entropy per unit rapidity
\begin{equation}
S= \int rdr d\phi  \  s(r,\phi ) \ .
\label{eq:S}
\end{equation}

\subsection{Transverse momentum -  harmonic flow correlations}

Event by event correlations between the average transverse momentum and the harmonic flow can be measured using the
correlation coefficient \cite{Bozek:2016yoj}
\begin{equation}
\rho(p_T,v_n^2)=\frac{Cov(p_T,v_n^2)}{\sqrt{Var(p_T) Var(v_n^2)}}
\label{eq:rhodef}
\end{equation}
with the covariance
\begin{equation}
Cov(p_T,v_n^2) =\langle p_T V_n V_n^\star \rangle - \langle p_T  \rangle \langle  V_n V_n^\star \rangle
\end{equation}
and the variances
\begin{equation}
Var(p_T) = \langle p_T^2   \rangle -  \langle p_T  \rangle^2 \ ,
\label{eq:varpt}
\end{equation}
\begin{equation}
Var(v_n^2) = \langle( V_n V_n^\star)^2   \rangle -  \langle V_n V_n^\star  \rangle^2 .
\label{eq:varvn}
\end{equation}
The covariance in the numerator of the correlation coefficient (\ref{eq:rhodef}) is a three particle correlators,
but the variance of the flow harmonic (\ref{eq:varvn}) is a four particle
correlator. The experimental estimators for covariance and the variances in (\ref{eq:rhodef}) involve up to  three or four   sums over particles in the event, with self-correlations
excluded \cite{Bozek:2016yoj}

\begin{figure}
\vspace{1mm}
\begin{center}
\includegraphics[width=0.48 \textwidth]{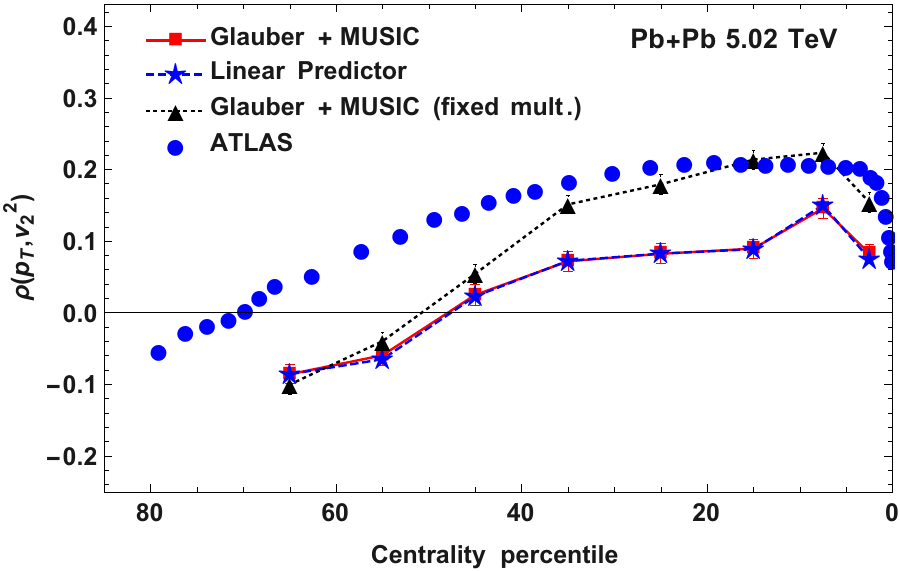} 
\end{center}
\caption{The correlation coefficient of the average transverse momentum and the elliptic flow coefficient in Pb+Pb collisions as a function of  centrality. The experimental data are from the ATLAS Collaboration \cite{ATLAS:2021kty} (blue points). The red squares denote the results of the hydrodynamic simulations, the black triangles  show the results for the correlation coefficient corrected for  multiplicity fluctuations (\ref{eq:partial})  and the  blue dashed line with stars shows   the correlation coefficient  obtained from the linear predictor (\ref{eq:lipred}). }
\label{fig:rho2}
\end{figure}

\begin{figure}
\vspace{1mm}
\begin{center}
\includegraphics[width=0.48 \textwidth]{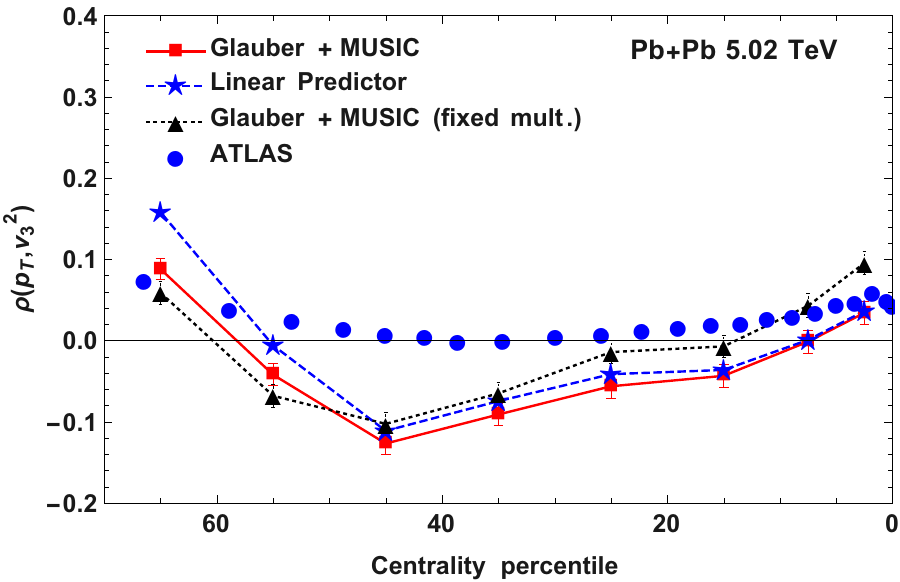} 
\end{center}
\caption{Same as in Fig. \ref{fig:rho2}, but for the triangular flow. }
\label{fig:rho3}
\end{figure}

The results for the transverse momentum-elliptic flow correlation coefficient $\rho(p_T,v_2^2)$ are shown in Fig. \ref{fig:rho2}.  The simulated correlation coefficients  present a qualitatively similar trend to the experimental data of the ATLAS Collaboration \cite{ATLAS:2021kty}. In particular, the correlation coefficient
$\rho(p_T,v_2^2)$ decreases in the most central collisions, and in peripheral collisions the correlation coefficient changes sign.

It should be noted that the change of sign of  $\rho(p_T,v_2^2)$ happens in the simulation for
more central collisions than in the data.
The measured correlation coefficient for  the triangular flow  $\rho(p_T,v_3^2)$ is small and is shown in Fig.\ref{fig:rho3}. The simulation does not fully describe this feature of the experimental data. The discrepancies observed between the data and simulations results for  $\rho(p_T,v_n^2)$ may
indicate that some physics of the dynamics in the model
or some essential correlations  in the initial state are missing. At very low
multiplicities some correlations may be due to non-flow contributions
and/or the initial
flow, but it is difficult to explain why a significant  discrepancy for   $\rho(p_T,v_2^2)$ is visible also for mid-central collisions.


The correlation between the average transverse momentum and the flow harmonics is partially due to correlations of those quantities with the event multiplicity. The experimental data is presented in narrow bins of centrality, where this effect is reduced. The dependence of the variance or covariance of studied observables on the fluctuations of a third variable, e.g. the multiplicity, can be taken into account by calculating the partial variance or covariance \cite{Olszewski:2017vyg}.
The partial correlation coefficient
\begin{equation}
\rho(p_T,v_n^2 \bullet N)=\frac{\rho(p_T,v_n^2)-\rho(p_T,N)\rho(N,v_n^2)}{\sqrt{1-\rho(p_T,N)^2}\sqrt{1-\rho(v_n^2,N)^2}}
\label{eq:partial}
\end{equation}
is an estimate of the correlation coefficient at fixed multiplicity  \cite{Bozek:2019wyr}.
The results for the partial correlation coefficients are shown in Figs. \ref{fig:rho2} and \ref{fig:rho3}  (black triangles). The correction is sizable for the
elliptic flow correlations coefficient $\rho(p_T,v_2^2)$. The partial correlation coefficient is closer to the experimental data.
Generally, if the dependence of an observable $O$, in a given bin,  on multiplicity is approximately linear, the
correction to $O$ can be implemented as \cite{Schenke:2020uqq}
\begin{equation}
\tilde{O}= O -\frac{Cov(O,N)}{Var(N)}\left(N-\langle N \rangle \right)
\label{eq:corrected}
\end{equation}
To correct for the effect of multiplicity fluctuation, we  can calculate event averages for moments
of the corrected variables (\ref{eq:corrected}).
For the correlation coefficient
this procedure is equivalent to the formula for the partial correlation coefficient (\ref{eq:partial}). In the following, we use  the corrected observables
$\tilde{O}$ to estimate higher order cumulants  without effects of
multiplicity fluctuations. When comparing to the experimental
data obtained
in narrow multiplicity bins the simulation results corrected for
multiplicity fluctuations should be considered. If the experimental
data are obtained in wide multiplicity bins or using a different
definition on centrality bins a correction similar to \ref{eq:corrected}
could be used in order to obtain a consistent definition of correlations and cumulants between different experiments and model calculations.

\subsection{Linear predictor}
The average transverse momentum and the harmonic flow coefficients are largely determined by the
initial conditions \cite{Gardim:2011xv,Qiu:2011iv,Niemi:2012aj,Broniowski:2009fm}. The eccentricities of the initial distribution
are strongly correlated with harmonic flow coefficients of the emitted particles.
The average transverse momentum of final particles can be predicted using the initial entropy $S$ (\ref{eq:S})
and the RMS size $R$ (\ref{eq:rms}) \cite{Mazeliauskas:2015efa,Bozek:2017elk}. Alternatively, a predictor of the final transverse momentum based
on the initial energy per rapidity  \cite{Giacalone:2020dln} or the energy weighted entropy  \cite{Schenke:2020uqq} can be used.

It has been noted that in order to describe the correlation of the transverse momentum with the elliptic or triangular flow,
the eccentricities  should be included in the predictor for the final transverse momentum \cite{Bozek:2020drh}.
It was shown that such an  improved predictor describes well the transverse momentum-harmonic flow correlation
coefficients, obtained from the full hydrodynamic simulation.
In the following we use a general  linear predictor, based on moments of the initial density
\begin{eqnarray}
\hat{v}_2^2 & =   & k_2 \epsilon_2^2 + \alpha_2 \delta R + \beta_2 \delta S \nonumber \\
\hat{v}_3^2 & =   & k_3 \epsilon_3^2 + \alpha_3 \delta R + \beta_3 \delta S\nonumber \\
\hat{p}_T & =   & \langle p_T\rangle + \alpha_p \delta R + \beta_p \delta S + \gamma_p \delta \epsilon_2^2 +
\lambda_p \delta \epsilon_3^2   \ \ \ ,
\label{eq:lipred}
\end{eqnarray}
where for any observable $\delta O =O -\langle O \rangle $.
The prediction of each of the observables in the above equations is optimized
separately. Only after fixing the parameters of the linear predictor, the cumulants between predicted observables are
calculated. The linear predictor can be written as
\begin{equation}
\delta \hat{O}_i= L^j_i \delta M_j
\end{equation}
with $M_i$ being the set of moments of the initial density. The moments of the final
observables can be written as a linear transformation from the moments in the initial state
\begin{equation}
\langle \delta O_i \dots \delta O_j \rangle = L^s_i \dots L^k_j \langle
\delta M_s \dots \delta M_k \rangle \ .
\end{equation}
The  linear predictor  (\ref{eq:lipred})
 describes fairly well the correlations and higher cumulants involving
 $p_T$, $v_2^2$, and $v_3^2$ (Compare blue stars (linear predictor) and red squares (hydrodynamic results) in Figs. \ref{fig:rho2}, \ref{fig:rho3}, \ref{fig:nscp23}, and \ref{fig:sscp23}). It demonstrates that the  proposed cumulants involving those observables can be understood as a linear hydrodynamic response
 on the correlations present in the initial state.

\section{Symmetric  cumulants }

\subsection{Second order cumulant}
The correlation between the magnitudes of the harmonic flows of different order has been studied using symmetric cumulants
\cite{Bilandzic:2013kga}.
 The second order symmetric cumulant is the covariance between the two observables
 \begin{equation}
 \langle A B \rangle -  \langle  A \rangle  \langle  B \rangle = Cov(A,B)  \ .
\end{equation}
The normalized symmetric cumulant is scaled by the averages of the observables in the covariance.
For the  transverse momentum and harmonic flow one may define
\begin{equation}
NSC(p_T,v_n^2)=\frac{\langle p_T v_n^2 \rangle - \langle p_T \rangle \langle v_n^2 \rangle}{\langle p_T \rangle \langle v_n^2 \rangle} \ .
\end{equation}



The information on the event by event correlations between the average transverse momentum and the harmonic flow is contained in the covariance of the two observables, and is basically the same in the correlation coefficient $\rho(p_T,v_n^2)$ and in the symmetric cumulant
$NSC(p_T,v_n^2)$.
The experimental extraction of the  normalized symmetric cumulant is simpler than for the correlation coefficient,
since the denominator involves at most a two particle correlator, where
methods to reduce non-flow effects can be implemented, even in small collision  systems \cite{Zhang:2021phk}.

\subsection{Third and fourth order cumulants}
Additional  information on correlations between flow observables is encoded in higher order cumulants
 \cite{Bilandzic:2013kga,Mordasini:2019hut,Moravcova:2020wnf}.
 The  $n$-th order cumulant involves only correlation
 of $n$ observables,
 with all lower order correlations subtracted.
 The third and fourth order symmetric cumulants for scalar observables have the form \cite{Mordasini:2019hut}
 \begin{eqnarray}
 SC(A,B,C) & = & \langle A B C  \rangle-\langle A B\rangle \langle C\rangle-\langle A C \rangle \langle B \rangle \nonumber \\ & & 
 -\langle B C \rangle \langle A \rangle+2 \langle A\rangle \langle B\rangle\langle C\rangle\label{eq:sc3}
 \end{eqnarray}
 and
 \begin{eqnarray}
 SC(A,B,C,D) & = & \langle A B C D \rangle-\langle A B C\rangle \langle D\rangle-\langle A B D \rangle \langle C \rangle \nonumber \\
 & & -\langle A C D \rangle \langle B \rangle-\langle B C D \rangle \langle A \rangle -\langle A B\rangle \langle C D\rangle \nonumber \\
& &  -\langle A C \rangle \langle B D\rangle \nonumber 
 -\langle B C \rangle \langle A D \rangle \nonumber \\
& &  +2\left( \langle A B\rangle \langle C\rangle \langle D\rangle +\langle A C\rangle \langle B\rangle \langle D\rangle \right. \nonumber \\
& & + \langle A D\rangle \langle C\rangle \langle B\rangle
 +\langle B C\rangle \langle A\rangle \langle D\rangle  \nonumber \\
 & & \left. +\langle B D \rangle \langle A\rangle \langle C\rangle +\langle C D \rangle \langle A\rangle \langle B\rangle \right) \nonumber \\
& &  -6 \langle A\rangle \langle B\rangle\langle C\rangle\langle D\rangle
 \label{eq:sc4}
 \end{eqnarray}
The corresponding normalized symmetric cumulants are
\begin{equation}
NSC(A,B,C)=\frac{SC(A,B,C)}{\langle A\rangle\langle B\rangle\langle C\rangle}
\label{eq:nsc3}
\end{equation}
and
\begin{equation}
NSC(A,B,C,D)=\frac{SC(A,B,C,D)}{\langle A\rangle\langle B\rangle\langle C\rangle\langle D\rangle}
\label{eq:nsc4}
\end{equation}
As noted, for higher order symmetric cumulants the effect of multiplicity fluctuations can be reduced using observables corrected for changes in multiplicity
(\ref{eq:corrected}).

\begin{figure}
\vspace{1mm}
\begin{center}
\includegraphics[width=0.48 \textwidth]{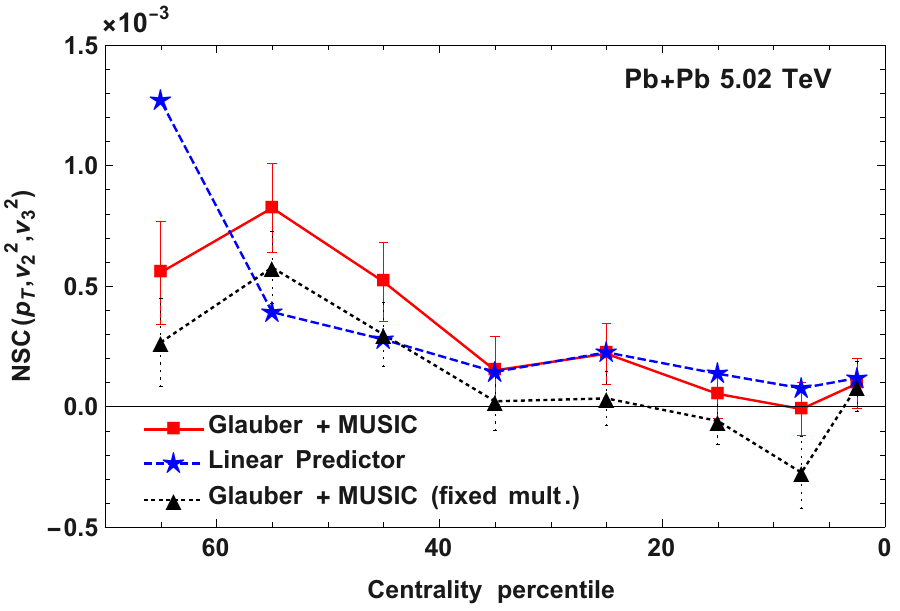} 
\end{center}
\caption{The normalized symmetric cumulant of the average transverse momentum and the squares of the elliptic and triangular flow (red squares).  The blue stars denote the results obtained from the linear predictor (\ref{eq:lipred}) and the black triangles show the scaled symmetric cumulant for observables corrected for multiplicity fluctuations (\ref{eq:corrected}). }
\label{fig:nscp23}
\end{figure}

\begin{figure}
\vspace{1mm}
\begin{center}
\includegraphics[width=0.48 \textwidth]{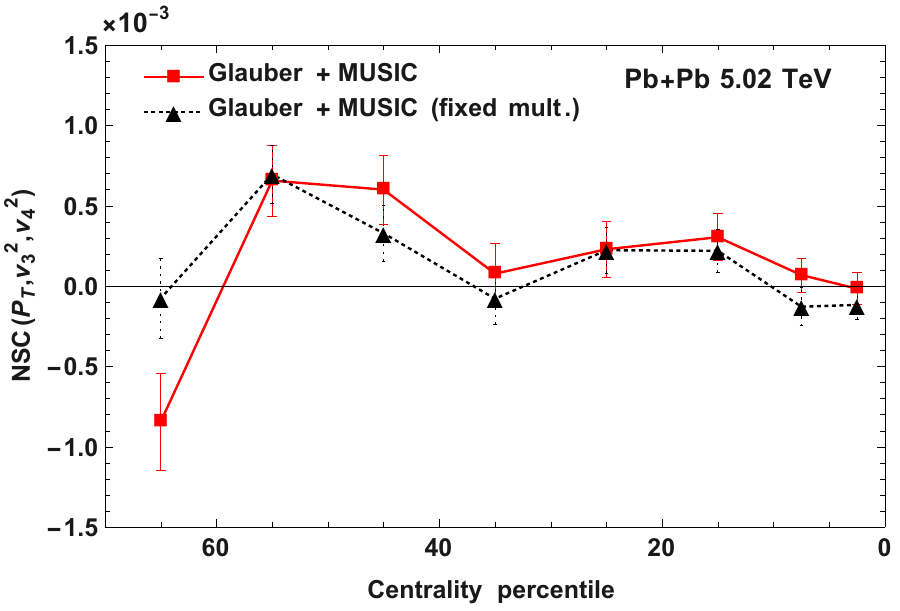} 
\end{center}
\caption{Same as in Fig. \ref{fig:nscp23} but for the triangular  and quadratic flows.}
\label{fig:nscp34}
\end{figure}

\begin{figure}
\vspace{1mm}
\begin{center}
\includegraphics[width=0.48 \textwidth]{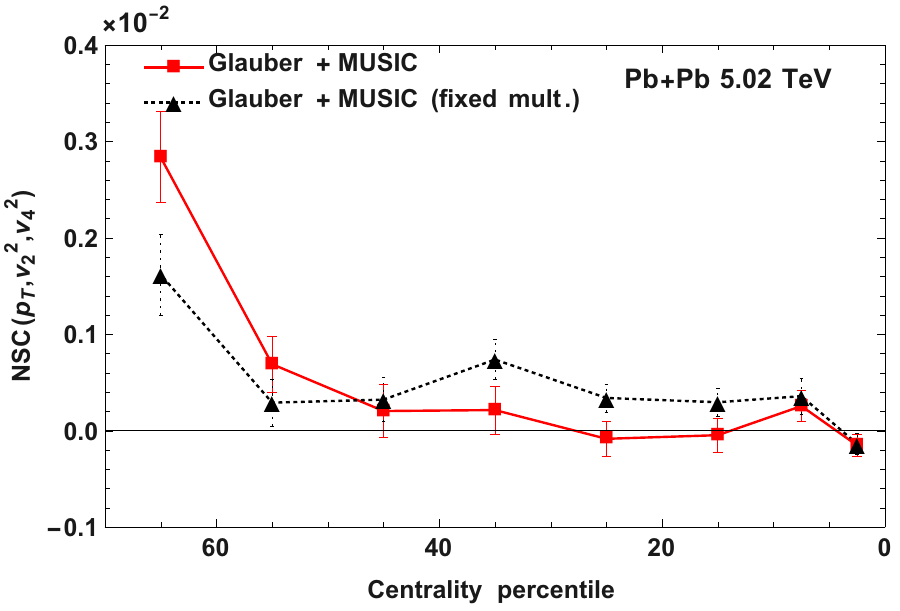} 
\end{center}
\caption{Same as in Fig. \ref{fig:nscp23} but for the elliptic and quadratic flows.}
\label{fig:nscp24}
\end{figure}


In Figs. \ref{fig:nscp23}, \ref{fig:nscp34}, and \ref{fig:nscp24}, the
third order cumulants for $p_T$ and two different coefficients of the harmonic flow are shown. For all the  third order correlations considered, the magnitude of the scaled cumulant is small, as it measures only genuine third order correlations. For the cumulants $NSC(p_T,v_2^2,v_3^2)$ and $NSC(p_T,v_2^2,v_4^2)$ an increase is visible in peripheral collisions. The linear predictor (\ref{eq:lipred}), based on initial correlations only,
describes the full hydrodynamic calculation for centralities 0-50\%. The cumulants involving $v_4$ cannot be predicted using a linear predictor, but could  serve as a precise measure of nonlinearities between harmonic flows of different order. The fourth order normalized symmetric cumulant $NSC(p_T,v_2^2,v_3^2,v_4^2)$ is compatible with zero within the  statistical
accuracy of our calculation  (not shown).

\section{Symmetric cumulants for collisions of   deformed nuclei}

In central collisions of deformed nuclei the relative orientation of the
axes of deformation of the two nuclei  determines the initial ellipticity,
entropy, and the size of the fireball
\cite{Heinz:2004ir,Goldschmidt:2015kpa,Rybczynski:2012av,Haque:2011aa,Chatterjee:2014sea,Adamczyk:2015obl,Giacalone:2020awm}. As a consequence, the final elliptic flow, the average transverse momentum and/or the  multiplicity may be correlated with the initial
orientation of the deformation axes in a collision.
In central collisions of deformed nuclei, tip-on-tip collision  lead to a
large transverse momentum and small elliptic flow, while body-on-body collisions
lead to a smaller transverse momentum but larger elliptic flow
\cite{Giacalone:2019pca}. This effect
reduces the value of the correlation coefficient $\rho(p_T,v_2^2)$ in central
U+U collisions.

\begin{figure}
\vspace{1mm}
\begin{center}
\includegraphics[width=0.48 \textwidth]{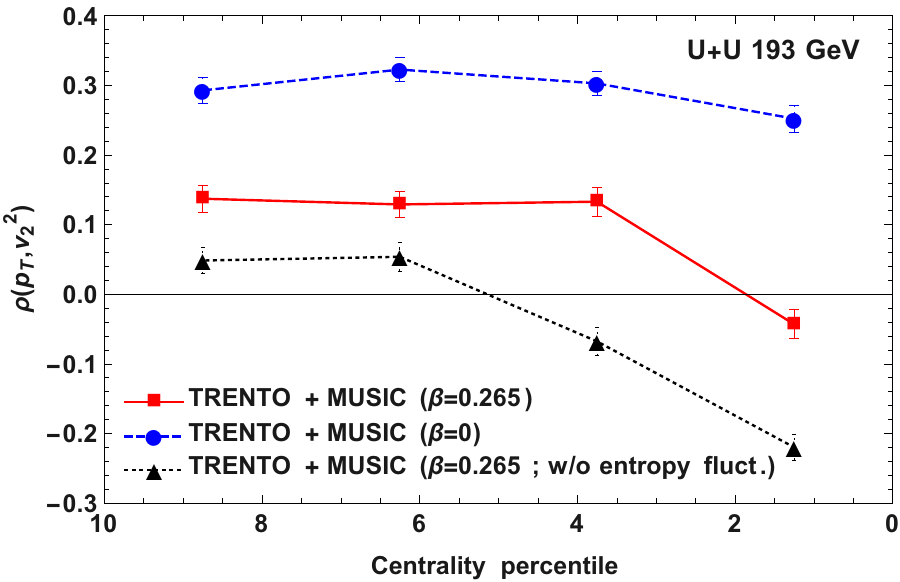} 
\end{center}
\caption{The correlation coefficient between the average transverse momentum and the elliptic flow coefficient $\rho(p_T,v_2^2)$ for central U+U collisions at
$\sqrt{s_{NN}}=193$~GeV as a function of centrality. Results for collisions of deformed nuclei with and without fluctuations in entropy deposition are denoted with red squares and black triangles respectively. The results for spherical nuclei are represented using blue dots. }
\label{fig:uurp2}
\end{figure}

\begin{figure}
\vspace{1mm}
\begin{center}
\includegraphics[width=0.48 \textwidth]{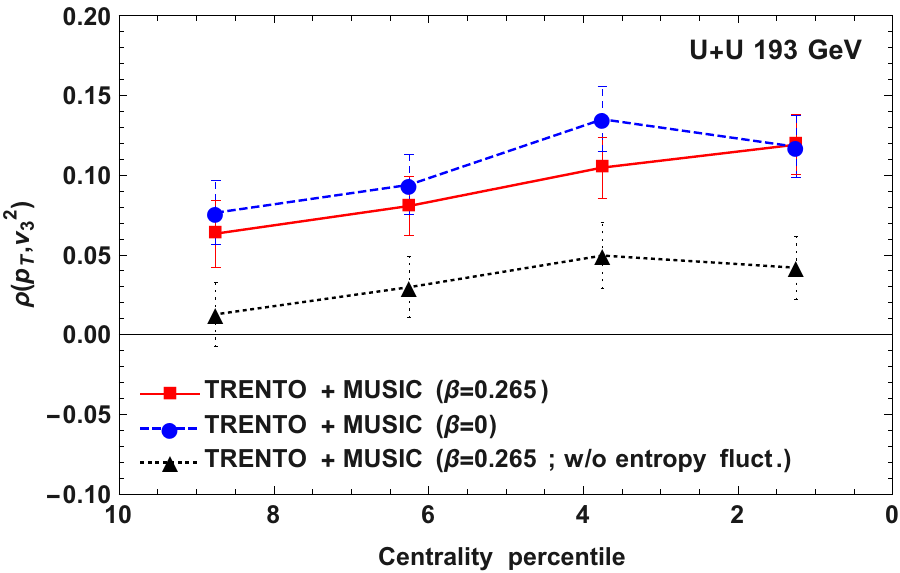} 
\end{center}
\caption{Same as in Fig. \ref{fig:uurp2} but for the triangular flow. }
\label{fig:uurp3}
\end{figure}

\begin{figure}
\vspace{1mm}
\begin{center}
\includegraphics[width=0.48 \textwidth]{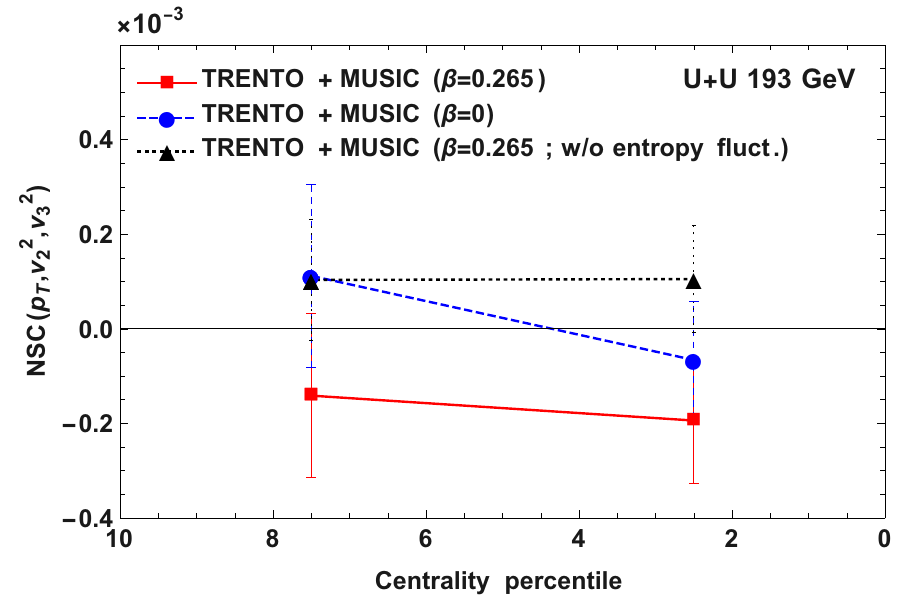} 
\end{center}
\caption{The normalized symmetric cumulant of  the average transverse
momentum,  the elliptic, and the triangular  flow coefficients
$NSC(p_T,v_2^2,v_3^2)$ for central U+U collisions at
$\sqrt{s_{NN}}=193$~GeV as function of centrality.  Results for collisions of deformed nuclei with and without fluctuations in entropy deposition are denoted with red squares and black triangles respectively. The results for spherical nuclei are represented using blue dots. }
\label{fig:uunscp23}
\end{figure}

In this section  symmetric cumulants for U+U collisions at
$\sqrt{s_{NN}}=193$~GeV are studied. Initial conditions from the
TRENTO model
(with parameter $p=0$) \cite{Moreland:2014oya} are evolved using
the 2-dimensional  version of the  MUSIC hydrodynamic code.
We compare three scenarios for the initial conditions~: collisions
of uranium nuclei with deformation parameter $\beta=0.265$ and
fluctuations of entropy deposition from each participant (exponential
distribution), collisions of spherical uranium nuclei ($\beta=0$),
and collisions of deformed uranium nuclei ($\beta=0.265$) without
fluctuations of entropy
deposition. In Figs. \ref{fig:uurp2} and \ref{fig:uurp3},
the correlation coefficients of the average transverse momentum and
harmonic flow are shown. The value of the correlation coefficient $\rho(p_T,v_2^2)$
is smaller for collisions of deformed nuclei than in the case of
spherical nuclei, as expected. We note that  the
correlation coefficient for initial conditions  without fluctuations
in entropy deposition is smaller than in the cases with fluctuations for
both correlation coefficients $\rho(p_T,v_2^2)$ and $\rho(p_T,v_3^2)$.
In central collisions, the
covariances of the harmonic  flow observables with  multiplicity are small. We have checked that for the centrality bins used in our study for U+U collisions ,  the corrections for multiplicity fluctuations  to the   correlation coefficients and higher order symmetric cumulants are small. In the following we show only the uncorrected  symmetric cumulants for U+U collisions.
The third order symmetric cumulants $NSC(p_T,v_2^2,v_3^2)$ is smaller for collisions of deformed nuclei with fluctuations of entropy deposition
than in the other two scenarios studied (Fig. \ref{fig:uunscp23}), but the effect is small.

\begin{figure}
\vspace{1mm}
\begin{center}
\includegraphics[width=0.48 \textwidth]{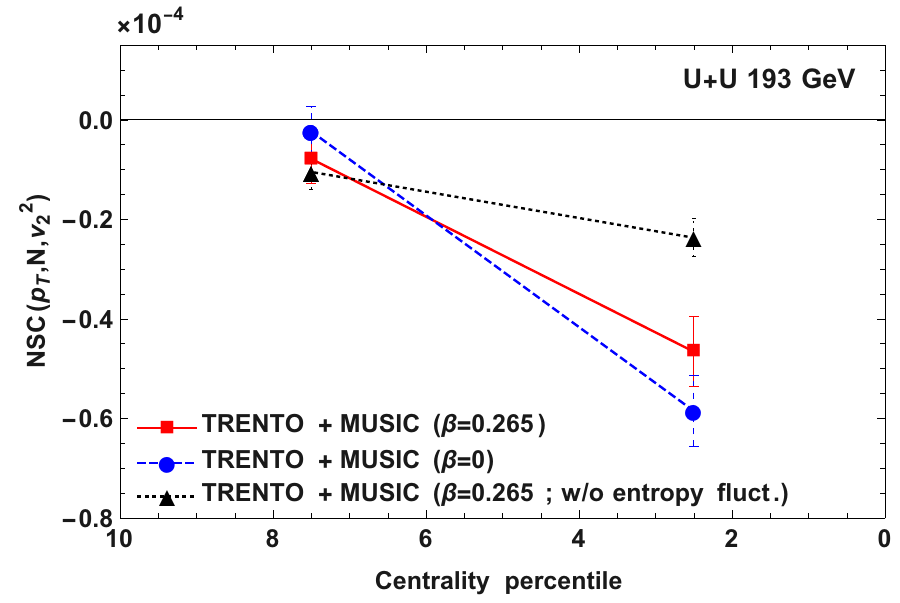} 
\end{center}
\caption{Same as in Fig. \ref{fig:uunscp23} but for the normalized symmetric cumulant of  the average transverse
momentum,  the multiplicity  and the elliptic flow coefficient
$NSC(p_T,N,v_2^2)$. Symbols as in Fig. \ref{fig:uunscp23}.}
\label{fig:uunscpn2}
\end{figure}

\begin{figure}
\vspace{1mm}
\begin{center}
\includegraphics[width=0.48 \textwidth]{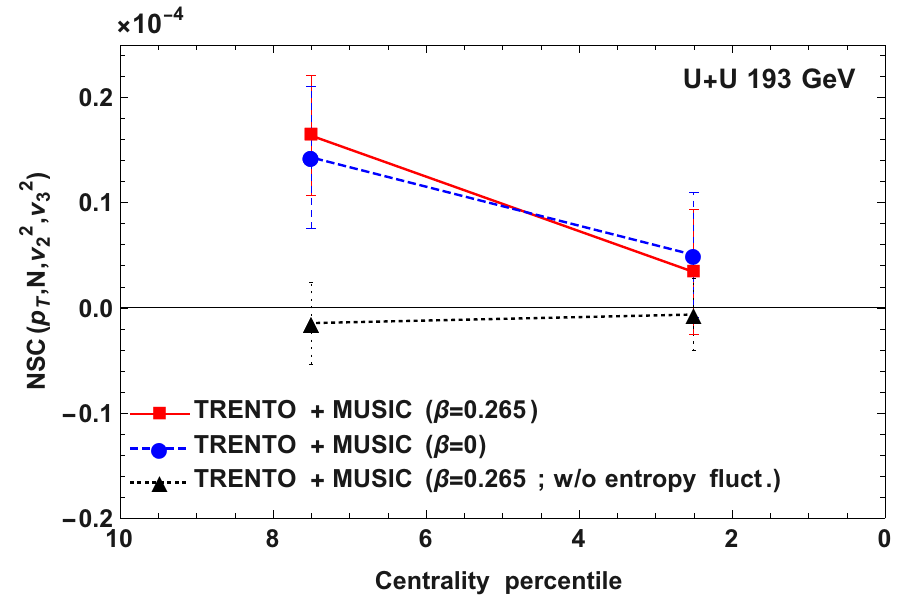} 
\end{center}
\caption{The fourth order  symmetric cumulant of the average transverse momentum, the multiplicity, the elliptic, and the triangular flow coefficients 
$NSC(p_T,N,v_2^2,v_3^2)$. Symbols as in Fig. \ref{fig:uunscp23}.}
\label{fig:uunscpn23}
\end{figure}

The collective flow  in central collisions of deformed nuclei is dominated by the fluctuations in the initial entropy and its  azimuthal asymmetries. Therefore,
it is interesting to study  symmetric cumulants involving not only the
average transverse momentum and harmonic flow, but also the multiplicity in
the event.
Please note, that the results for the  symmetric cumulants involving multiplicity as one of the observables may depend on the definition of the centrality bin. In order to reduce the bias from centrality cuts, that observables could be calculated experimentally using centrality bins based on forward rapidity observables, i.e. the forward transverse energy.
The symmetric cumulant $NSC(p_T,N,v_2^2)$ is sensitive to the fluctuations of entropy deposition from the participant nucleons (Fig. \ref{fig:uunscpn2}). Fluctuations in entropy deposition increase the magnitude of 
$NSC(p_T,N,v_2^2)$. The fourth order cumulant $NSC(p_T,N,v_2^2,v_3^2)$ is positive for simulations involving  fluctuations in the entropy deposition in the initial states (Fig. \ref{fig:uunscpn23}), while it is compatible with zero for
simulations without entropy fluctuations in the initial state. The normalized symmetric cumulant $NSC(p_T,v_2^2,v_3^2,v_4^2)$ (not shown) is compatible with zero within our statistical accuracy.


\section{Scaled symmetric cumulants}

\begin{figure}
\vspace{1mm}
\begin{center}
\includegraphics[width=0.48 \textwidth]{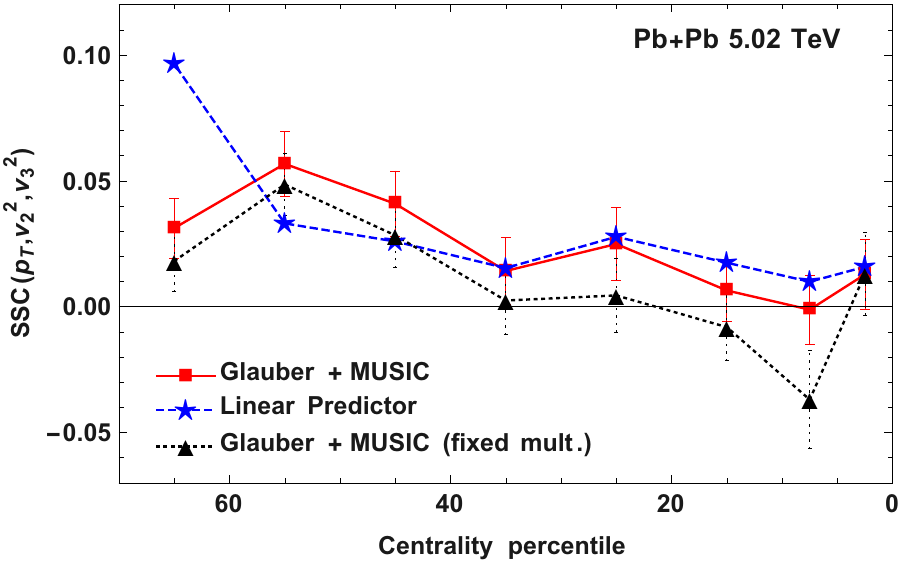} 
\end{center}
\caption{The scaled symmetric cumulant of the average transverse momentum and the squares of the elliptic and triangular flow coefficients $SSC(p_T,v_2^2,v_3^2)$ in Pb+Pb collisions as a function of centrality  (red squares).  The blue stars denote the results obtained from the linear predictor \ref{eq:lipred} and the black triangles show the scaled symmetric cumulant for observables corrected for multiplicity fluctuations \ref{eq:corrected}. }
\label{fig:sscp23}
\end{figure}

The normalized symmetric cumulants (\ref{eq:nsc3}) and (\ref{eq:nsc4})
involve, in the denominator, averages of the observables for which the cumulant is calculated.
However, the interpretation of the results is less obvious than for the correlation coefficient. Moreover, the average  transverse momentum in a collision depends on many effects \cite{Bozek:2012fw}, the freeze-out procedure, the bulk viscosity, the prequilibrium flow, or simply the experimental 
 transverse momentum range. The linear predictor for the average  transverse momentum in an event 
 (\ref{eq:lipred}) can predict  only deviations from the mean.

An alternative normalization of the symmetric cumulants would involve
the standard deviations of the observables in the denominator and we get the scaled symmetric cumulants,
\begin{equation}
SSC(A,B,C)=\frac{SC(A,B,C)}{\sqrt{Var(A)Var(B)Var(C)}}
\label{eq:ssc3}
\end{equation}
and
\begin{eqnarray}
&& SSC(A,B,C,D) =   \nonumber \\
&& \frac{SC(A,B,C,D)}{\sqrt{Var(A)Var(B)Var(C)Var(D)}} 
\label{eq:ssc4}
\end{eqnarray}
The scaled symmetric cumulant has the advantage that its prediction from the initial state does not require additional input on the value of the average transverse momentum (from simulation or experiment). The value of the scaled symmetric cumulants can be fully predicted using the linear hydrodynamic response.

\begin{figure}
\vspace{1mm}
\begin{center}
\includegraphics[width=0.48 \textwidth]{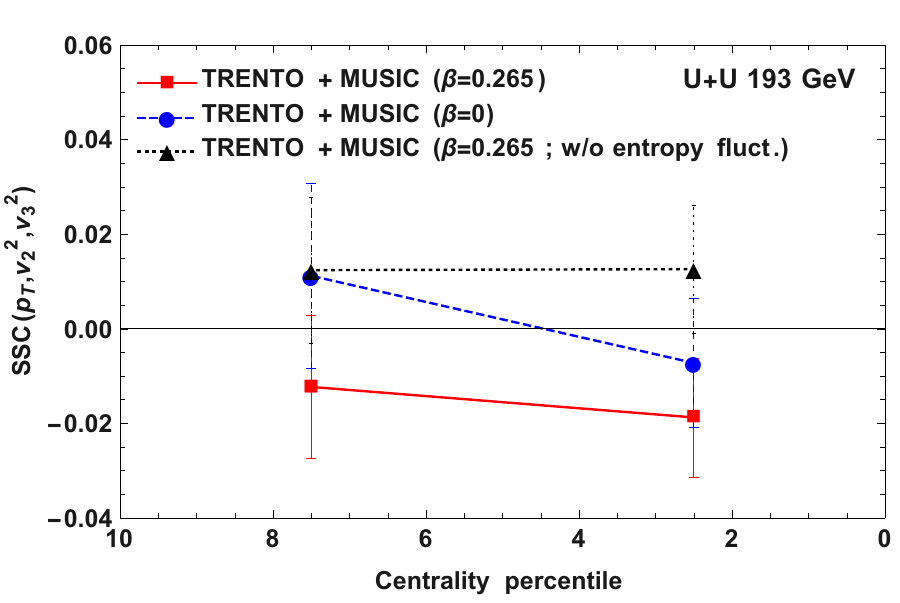} 
\end{center}
\caption{The scaled symmetric cumulant of  the average transverse
momentum,  the elliptic, and the triangular  flow coefficients
$SSC(p_T,v_2^2,v_3^2)$ for central U+U collisions at
$\sqrt{s_{NN}}=193$~GeV as a function of centrality.  Results for collisions of deformed nuclei with and without fluctuations in entropy deposition are denoted with red squares and black triangles respectively. The results for spherical nuclei are represented using blue dots. }
\label{fig:uusscp23}
\end{figure}

The scaled symmetric cumulants show a very similar behavior as the normalized symmetric cumulants discussed in previous sections. This can be seen in the examples chosen for Pb+Pb collisions (Fig. \ref{fig:sscp23}) and U+U collisions
(Fig. \ref{fig:uusscp23}). The numerical values are larger for the scaled symmetric cumulants than for the normalized symmetric cumulants. Alone the change from the normalization by the average transverse momentum to the normalization by its 
standard deviation  yields a factor in the range: $20 - 100$.


\section{Summary}

Flow observables are fluctuating from event to event. Moments
from the values of  several  flow observables in the event can be 
constructed. We propose to measure higher order
symmetric cumulants between the average transverse momentum and harmonic
flow coefficients. Symmetric  cumulants of order $n$ measure genuine $n$
order correlations between the  observables studied. 

Symmetric cumulants can be normalized by the averages or by the standard deviations of the considered observables. Such normalized 
cumulants involving the elliptic flow, the triangular flow,
and the transverse momentum could be predicted from the initial conditions.
Their study
could serve as a sensitive probe of higher order correlations in the
initial state of the evolution in the collision. Further cumulants involving
flow harmonics of higher order could serve as a probe of nonlinearities
in the hydrodynamic response.

We present predictions based on the hydrodynamic model for
collisions of spherical (Pb+Pb) or deformed (U+U) nuclei. The symmetric
cumulants involving the elliptic and triangular flows  obtained from  the full
hydrodynamic  calculation can be well described using a linear predictor
from the initial state. Our predictions could be compared to future experimental
measurements.

\section*{Acknowledgments}

This research is supported by the AGH University of Science and Technology and  by the  Polish National Science Centre grant 2019/35/O/ST2/00357.
\bibliography{../hydr}

\begin{thebibliography}{10}%
\makeatletter
\providecommand \@ifxundefined [1]{%
 \ifx #1\undefined \expandafter \@firstoftwo
 \else \expandafter \@secondoftwo
\fi
}%
\providecommand \@ifnum [1]{%
 \ifnum #1\expandafter \@firstoftwo
 \else \expandafter \@secondoftwo
\fi
}%
\providecommand \enquote [1]{``#1''}%
\providecommand \bibnamefont  [1]{#1}%
\providecommand \bibfnamefont [1]{#1}%
\providecommand \citenamefont [1]{#1}%
\providecommand\href[0]{\@sanitize\@href}%
\providecommand\@href[1]{\endgroup\@@startlink{#1}\endgroup\@@href}%
\providecommand\@@href[1]{#1\@@endlink}%
\providecommand \@sanitize [0]{\begingroup\catcode`\&12\catcode`\#12\relax}%
\@ifxundefined \pdfoutput {\@firstoftwo}{%
 \@ifnum{\z@=\pdfoutput}{\@firstoftwo}{\@secondoftwo}%
}{%
 \providecommand\@@startlink[1]{\leavevmode\special{html:<a href="#1">}}%
 \providecommand\@@endlink[0]{\special{html:</a>}}%
}{%
 \providecommand\@@startlink[1]{%
  \leavevmode
  \pdfstartlink
   attr{/Border[0 0 1 ]/H/I/C[0 1 1]}%
   user{/Subtype/Link/A<</Type/Action/S/URI/URI(#1)>>}%
  \relax
 }%
 \providecommand\@@endlink[0]{\pdfendlink}%
}%
\providecommand \url  [0]{\begingroup\@sanitize \@url }%
\providecommand \@url [1]{\endgroup\@href {#1}{\urlprefix}}%
\providecommand \urlprefix [0]{URL }%
\providecommand \Eprint[0]{\href }%
\@ifxundefined \urlstyle {%
  \providecommand \doi [1]{doi:\discretionary{}{}{}#1}%
}{%
  \providecommand \doi [0]{doi:\discretionary{}{}{}\begingroup
  \urlstyle{rm}\Url }%
}%
\providecommand \doibase [0]{http://dx.doi.org/}%
\providecommand \Doi[1]{\href{\doibase#1}}%
\providecommand \bibAnnote [3]{%
  \BibitemShut{#1}%
  \begin{quotation}\noindent
    \textsc{Key:}\ #2\\\textsc{Annotation:}\ #3%
  \end{quotation}%
}%
\providecommand \bibAnnoteFile [2]{%
  \IfFileExists{#2}{\bibAnnote {#1} {#2} {\input{#2}}}{}%
}%
\providecommand \typeout [0]{\immediate \write \m@ne }%
\providecommand \selectlanguage [0]{\@gobble}%
\providecommand \bibinfo [0]{\@secondoftwo}%
\providecommand \bibfield [0]{\@secondoftwo}%
\providecommand \translation [1]{[#1]}%
\providecommand \BibitemOpen[0]{}%
\providecommand \bibitemStop [0]{}%
\providecommand \bibitemNoStop [0]{.\EOS\space}%
\providecommand \EOS [0]{\spacefactor3000\relax}%
\providecommand \BibitemShut [1]{\csname bibitem#1\endcsname}%
\bibitem{Ollitrault:2010tn}%
  \BibitemOpen
  \bibfield{author}{%
  \bibinfo {author} {\bibfnamefont{J.-Y.}\ \bibnamefont{Ollitrault}},\ }%
  \bibfield{journal}{%
  \Doi{10.1088/1742-6596/312/1/012002}{\bibinfo {journal} {J. Phys. Conf.
  Ser.}}\ }%
  \textbf{\bibinfo {volume} {312}},\ \bibinfo {pages} {012002} (\bibinfo {year}
  {2011})%
  \bibAnnoteFile{NoStop}{Ollitrault:2010tn}%
\bibitem{Heinz:2013th}%
  \BibitemOpen
  \bibfield{author}{%
  \bibinfo {author} {\bibfnamefont{U.}~\bibnamefont{Heinz}}\ and\ \bibinfo
  {author} {\bibfnamefont{R.}~\bibnamefont{Snellings}},\ }%
  \bibfield{journal}{%
  \Doi{10.1146/annurev-nucl-102212-170540}{\bibinfo {journal}
  {Ann.Rev.Nucl.Part.Sci.}}\ }%
  \textbf{\bibinfo {volume} {63}},\ \bibinfo {pages} {123} (\bibinfo {year}
  {2013})%
  \bibAnnoteFile{NoStop}{Heinz:2013th}%
\bibitem{Gale:2013da}%
  \BibitemOpen
  \bibfield{author}{%
  \bibinfo {author} {\bibfnamefont{C.}~\bibnamefont{Gale}}, \bibinfo {author}
  {\bibfnamefont{S.}~\bibnamefont{Jeon}},\ and\ \bibinfo {author}
  {\bibfnamefont{B.}~\bibnamefont{Schenke}},\ }%
  \bibfield{journal}{%
  \Doi{10.1142/S0217751X13400113}{\bibinfo {journal} {Int.J.Mod.Phys.}}\ }%
  \textbf{\bibinfo {volume} {A28}},\ \bibinfo {pages} {1340011} (\bibinfo
  {year} {2013})%
  \bibAnnoteFile{NoStop}{Gale:2013da}%
\bibitem{Broniowski:2009fm}%
  \BibitemOpen
  \bibfield{author}{%
  \bibinfo {author} {\bibfnamefont{W.}~\bibnamefont{Broniowski}}, \bibinfo
  {author} {\bibfnamefont{M.}~\bibnamefont{Chojnacki}},\ and\ \bibinfo {author}
  {\bibfnamefont{L.}~\bibnamefont{Obara}},\ }%
  \bibfield{journal}{%
  \Doi{10.1103/PhysRevC.80.051902}{\bibinfo {journal} {Phys. Rev.}}\ }%
  \textbf{\bibinfo {volume} {C80}},\ \bibinfo {pages} {051902} (\bibinfo {year}
  {2009})%
  \bibAnnoteFile{NoStop}{Broniowski:2009fm}%
\bibitem{Bozek:2016yoj}%
  \BibitemOpen
  \bibfield{author}{%
  \bibinfo {author} {\bibfnamefont{P.}~\bibnamefont{Bo{\.z}ek}},\ }%
  \bibfield{journal}{%
  \Doi{10.1103/PhysRevC.93.044908}{\bibinfo {journal} {Phys. Rev.}}\ }%
  \textbf{\bibinfo {volume} {C93}},\ \bibinfo {pages} {044908} (\bibinfo {year}
  {2016})%
  \bibAnnoteFile{NoStop}{Bozek:2016yoj}%
\bibitem{Bozek:2020drh}%
  \BibitemOpen
  \bibfield{author}{%
  \bibinfo {author} {\bibfnamefont{P.}~\bibnamefont{Bo{\.z}ek}}\ and\ \bibinfo
  {author} {\bibfnamefont{H.}~\bibnamefont{Mehrabpour}},\ }%
  \bibfield{journal}{%
  \Doi{10.1103/PhysRevC.101.064902}{\bibinfo {journal} {Phys. Rev. C}}\ }%
  \textbf{\bibinfo {volume} {101}},\ \bibinfo {pages} {064902} (\bibinfo {year}
  {2020})%
  \bibAnnoteFile{NoStop}{Bozek:2020drh}%
\bibitem{Schenke:2020uqq}%
  \BibitemOpen
  \bibfield{author}{%
  \bibinfo {author} {\bibfnamefont{B.}~\bibnamefont{Schenke}}, \bibinfo
  {author} {\bibfnamefont{C.}~\bibnamefont{Shen}},\ and\ \bibinfo {author}
  {\bibfnamefont{D.}~\bibnamefont{Teaney}},\ }%
  \bibfield{journal}{%
  \Doi{10.1103/PhysRevC.102.034905}{\bibinfo {journal} {Phys. Rev. C}}\ }%
  \textbf{\bibinfo {volume} {102}},\ \bibinfo {pages} {034905} (\bibinfo {year}
  {2020})%
  \bibAnnoteFile{NoStop}{Schenke:2020uqq}%
\bibitem{Giacalone:2020dln}%
  \BibitemOpen
  \bibfield{author}{%
  \bibinfo {author} {\bibfnamefont{G.}~\bibnamefont{Giacalone}}, \bibinfo
  {author} {\bibfnamefont{F.~G.}\ \bibnamefont{Gardim}}, \bibinfo {author}
  {\bibfnamefont{J.}~\bibnamefont{Noronha-Hostler}},\ and\ \bibinfo {author}
  {\bibfnamefont{J.-Y.}\ \bibnamefont{Ollitrault}},\ }%
  \bibfield{journal}{%
  \Doi{10.1103/PhysRevC.103.024909}{\bibinfo {journal} {Phys. Rev. C}}\ }%
  \textbf{\bibinfo {volume} {103}},\ \bibinfo {pages} {024909} (\bibinfo {year}
  {2021})%
  \bibAnnoteFile{NoStop}{Giacalone:2020dln}%
\bibitem{Aad:2019fgl}%
  \BibitemOpen
  \bibfield{author}{%
  \bibinfo {author} {\bibfnamefont{G.}~\bibnamefont{Aad}} \emph{et~al.}
  (\bibinfo {collaboration} {ATLAS}),\ }%
  \bibfield{journal}{%
  \Doi{10.1140/epjc/s10052-019-7489-6}{\bibinfo {journal} {Eur. Phys. J.}}\ }%
  \textbf{\bibinfo {volume} {C79}},\ \bibinfo {pages} {985} (\bibinfo {year}
  {2019})%
  \bibAnnoteFile{NoStop}{Aad:2019fgl}%
\bibitem{Giacalone:2019pca}%
  \BibitemOpen
  \bibfield{author}{%
  \bibinfo {author} {\bibfnamefont{G.}~\bibnamefont{Giacalone}},\ }%
  \bibfield{journal}{%
  \Doi{10.1103/PhysRevLett.124.202301}{\bibinfo {journal} {Phys. Rev. Lett.}}\
  }%
  \textbf{\bibinfo {volume} {124}},\ \bibinfo {pages} {202301} (\bibinfo {year}
  {2020})%
  \bibAnnoteFile{NoStop}{Giacalone:2019pca}%
\bibitem{Giacalone:2020awm}%
  \BibitemOpen
  \bibfield{author}{%
  \bibinfo {author} {\bibfnamefont{G.}~\bibnamefont{Giacalone}},\ }%
  \bibfield{journal}{%
  \Doi{10.1103/PhysRevC.102.024901}{\bibinfo {journal} {Phys. Rev. C}}\ }%
  \textbf{\bibinfo {volume} {102}},\ \bibinfo {pages} {024901} (\bibinfo {year}
  {2020})%
  \bibAnnoteFile{NoStop}{Giacalone:2020awm}%
\bibitem{ATLAS:2021kty}%
  \BibitemOpen
  \bibfield{author}{%
  \bibinfo {author} {\bibfnamefont{G.}~\bibnamefont{Aad}} \emph{et~al.}
  (\bibinfo {collaboration} {ATLAS}),\ }%
  \bibfield{journal}{%
  \bibinfo {journal} {ATLAS-CONF-2021-001}}%
   (\bibinfo {year} {2021})%
  \bibAnnoteFile{NoStop}{ATLAS:2021kty}%
\bibitem{JiaIS2021}%
  \BibitemOpen
  \bibfield{author}{%
  \bibinfo {author} {\bibfnamefont{J.}~\bibnamefont{Jia}} (\bibinfo
  {collaboration} {STAR}),\ }%
  \bibfield{journal}{%
  \bibinfo {journal} {talk presented at 6th International Conference on the
  Initial Stages in High-Energy Nuclear Collisions. Revhovot, 10-15 January~}}%
   (\bibinfo {year} {2021})%
  \bibAnnoteFile{NoStop}{JiaIS2021}%
\bibitem{Giacalone:2020byk}%
  \BibitemOpen
  \bibfield{author}{%
  \bibinfo {author} {\bibfnamefont{G.}~\bibnamefont{Giacalone}}, \bibinfo
  {author} {\bibfnamefont{B.}~\bibnamefont{Schenke}},\ and\ \bibinfo {author}
  {\bibfnamefont{C.}~\bibnamefont{Shen}},\ }%
  \bibfield{journal}{%
  \Doi{10.1103/PhysRevLett.125.192301}{\bibinfo {journal} {Phys. Rev. Lett.}}\
  }%
  \textbf{\bibinfo {volume} {125}},\ \bibinfo {pages} {192301} (\bibinfo {year}
  {2020})%
  \bibAnnoteFile{NoStop}{Giacalone:2020byk}%
\bibitem{Bilandzic:2013kga}%
  \BibitemOpen
  \bibfield{author}{%
  \bibinfo {author} {\bibfnamefont{A.}~\bibnamefont{Bilandzic}}, \bibinfo
  {author} {\bibfnamefont{C.~H.}\ \bibnamefont{Christensen}}, \bibinfo {author}
  {\bibfnamefont{K.}~\bibnamefont{Gulbrandsen}}, \bibinfo {author}
  {\bibfnamefont{A.}~\bibnamefont{Hansen}},\ and\ \bibinfo {author}
  {\bibfnamefont{Y.}~\bibnamefont{Zhou}},\ }%
  \bibfield{journal}{%
  \Doi{10.1103/PhysRevC.89.064904}{\bibinfo {journal} {Phys. Rev. C}}\ }%
  \textbf{\bibinfo {volume} {89}},\ \bibinfo {pages} {064904} (\bibinfo {year}
  {2014})%
  \bibAnnoteFile{NoStop}{Bilandzic:2013kga}%
\bibitem{Mordasini:2019hut}%
  \BibitemOpen
  \bibfield{author}{%
  \bibinfo {author} {\bibfnamefont{C.}~\bibnamefont{Mordasini}}, \bibinfo
  {author} {\bibfnamefont{A.}~\bibnamefont{Bilandzic}}, \bibinfo {author}
  {\bibfnamefont{D.}~\bibnamefont{Karako\c{c}}},\ and\ \bibinfo {author}
  {\bibfnamefont{S.~F.}\ \bibnamefont{Taghavi}},\ }%
  \bibfield{journal}{%
  \Doi{10.1103/PhysRevC.102.024907}{\bibinfo {journal} {Phys. Rev. C}}\ }%
  \textbf{\bibinfo {volume} {102}},\ \bibinfo {pages} {024907} (\bibinfo {year}
  {2020})%
  \bibAnnoteFile{NoStop}{Mordasini:2019hut}%
\bibitem{Moravcova:2020wnf}%
  \BibitemOpen
  \bibfield{author}{%
  \bibinfo {author} {\bibfnamefont{Z.}~\bibnamefont{Moravcova}}, \bibinfo
  {author} {\bibfnamefont{K.}~\bibnamefont{Gulbrandsen}},\ and\ \bibinfo
  {author} {\bibfnamefont{Y.}~\bibnamefont{Zhou}}}%
   (\bibinfo {month} {5}\ \bibinfo {year} {2020}),\
  \Eprint{http://arxiv.org/abs/2005.07974}{arXiv:2005.07974 [nucl-th]}%
  \bibAnnoteFile{NoStop}{Moravcova:2020wnf}%
\bibitem{Schenke:2010nt}%
  \BibitemOpen
  \bibfield{author}{%
  \bibinfo {author} {\bibfnamefont{B.}~\bibnamefont{Schenke}}, \bibinfo
  {author} {\bibfnamefont{S.}~\bibnamefont{Jeon}},\ and\ \bibinfo {author}
  {\bibfnamefont{C.}~\bibnamefont{Gale}},\ }%
  \bibfield{journal}{%
  \Doi{10.1103/PhysRevC.82.014903}{\bibinfo {journal} {Phys. Rev. C}}\ }%
  \textbf{\bibinfo {volume} {82}},\ \bibinfo {pages} {014903} (\bibinfo {year}
  {2010})%
  \bibAnnoteFile{NoStop}{Schenke:2010nt}%
\bibitem{Schenke:2010rr}%
  \BibitemOpen
  \bibfield{author}{%
  \bibinfo {author} {\bibfnamefont{B.}~\bibnamefont{Schenke}}, \bibinfo
  {author} {\bibfnamefont{S.}~\bibnamefont{Jeon}},\ and\ \bibinfo {author}
  {\bibfnamefont{C.}~\bibnamefont{Gale}},\ }%
  \bibfield{journal}{%
  \Doi{10.1103/PhysRevLett.106.042301}{\bibinfo {journal} {Phys. Rev. Lett.}}\
  }%
  \textbf{\bibinfo {volume} {106}},\ \bibinfo {pages} {042301} (\bibinfo {year}
  {2011})%
  \bibAnnoteFile{NoStop}{Schenke:2010rr}%
\bibitem{Paquet:2015lta}%
  \BibitemOpen
  \bibfield{author}{%
  \bibinfo {author} {\bibfnamefont{J.-F.}\ \bibnamefont{Paquet}}, \bibinfo
  {author} {\bibfnamefont{C.}~\bibnamefont{Shen}}, \bibinfo {author}
  {\bibfnamefont{G.~S.}\ \bibnamefont{Denicol}}, \bibinfo {author}
  {\bibfnamefont{M.}~\bibnamefont{Luzum}}, \bibinfo {author}
  {\bibfnamefont{B.}~\bibnamefont{Schenke}}, \bibinfo {author}
  {\bibfnamefont{S.}~\bibnamefont{Jeon}},\ and\ \bibinfo {author}
  {\bibfnamefont{C.}~\bibnamefont{Gale}},\ }%
  \bibfield{journal}{%
  \Doi{10.1103/PhysRevC.93.044906}{\bibinfo {journal} {Phys. Rev. C}}\ }%
  \textbf{\bibinfo {volume} {93}},\ \bibinfo {pages} {044906} (\bibinfo {year}
  {2016})%
  \bibAnnoteFile{NoStop}{Paquet:2015lta}%
\bibitem{Bozek:2019wyr}%
  \BibitemOpen
  \bibfield{author}{%
  \bibinfo {author} {\bibfnamefont{P.}~\bibnamefont{Bo\.zek}}, \bibinfo
  {author} {\bibfnamefont{W.}~\bibnamefont{Broniowski}}, \bibinfo {author}
  {\bibfnamefont{M.}~\bibnamefont{Rybczynski}},\ and\ \bibinfo {author}
  {\bibfnamefont{G.}~\bibnamefont{Stefanek}},\ }%
  \bibfield{journal}{%
  \Doi{10.1016/j.cpc.2019.07.014}{\bibinfo {journal} {Comput. Phys. Commun.}}\
  }%
  \textbf{\bibinfo {volume} {245}},\ \bibinfo {pages} {106850} (\bibinfo {year}
  {2019})%
  \bibAnnoteFile{NoStop}{Bozek:2019wyr}%
\bibitem{Olszewski:2017vyg}%
  \BibitemOpen
  \bibfield{author}{%
  \bibinfo {author} {\bibfnamefont{A.}~\bibnamefont{Olszewski}}\ and\ \bibinfo
  {author} {\bibfnamefont{W.}~\bibnamefont{Broniowski}},\ }%
  \bibfield{journal}{%
  \Doi{10.1103/PhysRevC.96.054903}{\bibinfo {journal} {Phys. Rev.}}\ }%
  \textbf{\bibinfo {volume} {C96}},\ \bibinfo {pages} {054903} (\bibinfo {year}
  {2017})%
  \bibAnnoteFile{NoStop}{Olszewski:2017vyg}%
\bibitem{Gardim:2011xv}%
  \BibitemOpen
  \bibfield{author}{%
  \bibinfo {author} {\bibfnamefont{F.~G.}\ \bibnamefont{Gardim}}, \bibinfo
  {author} {\bibfnamefont{F.}~\bibnamefont{Grassi}}, \bibinfo {author}
  {\bibfnamefont{M.}~\bibnamefont{Luzum}},\ and\ \bibinfo {author}
  {\bibfnamefont{J.-Y.}\ \bibnamefont{Ollitrault}},\ }%
  \bibfield{journal}{%
  \Doi{10.1103/PhysRevC.85.024908}{\bibinfo {journal} {Phys. Rev.}}\ }%
  \textbf{\bibinfo {volume} {C85}},\ \bibinfo {pages} {024908} (\bibinfo {year}
  {2012})%
  \bibAnnoteFile{NoStop}{Gardim:2011xv}%
\bibitem{Qiu:2011iv}%
  \BibitemOpen
  \bibfield{author}{%
  \bibinfo {author} {\bibfnamefont{Z.}~\bibnamefont{Qiu}}\ and\ \bibinfo
  {author} {\bibfnamefont{U.~W.}\ \bibnamefont{Heinz}},\ }%
  \bibfield{journal}{%
  \Doi{10.1103/PhysRevC.84.024911}{\bibinfo {journal} {Phys. Rev.}}\ }%
  \textbf{\bibinfo {volume} {C84}},\ \bibinfo {pages} {024911} (\bibinfo {year}
  {2011})%
  \bibAnnoteFile{NoStop}{Qiu:2011iv}%
\bibitem{Niemi:2012aj}%
  \BibitemOpen
  \bibfield{author}{%
  \bibinfo {author} {\bibfnamefont{H.}~\bibnamefont{Niemi}}, \bibinfo {author}
  {\bibfnamefont{G.~S.}\ \bibnamefont{Denicol}}, \bibinfo {author}
  {\bibfnamefont{H.}~\bibnamefont{Holopainen}},\ and\ \bibinfo {author}
  {\bibfnamefont{P.}~\bibnamefont{Huovinen}},\ }%
  \bibfield{journal}{%
  \Doi{10.1103/PhysRevC.87.054901}{\bibinfo {journal} {Phys. Rev.}}\ }%
  \textbf{\bibinfo {volume} {C87}},\ \bibinfo {pages} {054901} (\bibinfo {year}
  {2013})%
  \bibAnnoteFile{NoStop}{Niemi:2012aj}%
\bibitem{Mazeliauskas:2015efa}%
  \BibitemOpen
  \bibfield{author}{%
  \bibinfo {author} {\bibfnamefont{A.}~\bibnamefont{Mazeliauskas}}\ and\
  \bibinfo {author} {\bibfnamefont{D.}~\bibnamefont{Teaney}},\ }%
  \bibfield{journal}{%
  \Doi{10.1103/PhysRevC.93.024913}{\bibinfo {journal} {Phys. Rev.}}\ }%
  \textbf{\bibinfo {volume} {C93}},\ \bibinfo {pages} {024913} (\bibinfo {year}
  {2016})%
  \bibAnnoteFile{NoStop}{Mazeliauskas:2015efa}%
\bibitem{Bozek:2017elk}%
  \BibitemOpen
  \bibfield{author}{%
  \bibinfo {author} {\bibfnamefont{P.}~\bibnamefont{Bo{\.z}ek}}\ and\ \bibinfo
  {author} {\bibfnamefont{W.}~\bibnamefont{Broniowski}},\ }%
  \bibfield{journal}{%
  \Doi{10.1103/PhysRevC.96.014904}{\bibinfo {journal} {Phys. Rev.}}\ }%
  \textbf{\bibinfo {volume} {C96}},\ \bibinfo {pages} {014904} (\bibinfo {year}
  {2017})%
  \bibAnnoteFile{NoStop}{Bozek:2017elk}%
\bibitem{Zhang:2021phk}%
  \BibitemOpen
  \bibfield{author}{%
  \bibinfo {author} {\bibfnamefont{C.}~\bibnamefont{Zhang}}, \bibinfo {author}
  {\bibfnamefont{A.}~\bibnamefont{Behera}}, \bibinfo {author}
  {\bibfnamefont{S.}~\bibnamefont{Bhatta}},\ and\ \bibinfo {author}
  {\bibfnamefont{J.}~\bibnamefont{Jia}}}%
   (\bibinfo {year} {2021}),\
  \Eprint{http://arxiv.org/abs/2102.05200}{arXiv:2102.05200 [nucl-th]}%
  \bibAnnoteFile{NoStop}{Zhang:2021phk}%
\bibitem{Heinz:2004ir}%
  \BibitemOpen
  \bibfield{author}{%
  \bibinfo {author} {\bibfnamefont{U.~W.}\ \bibnamefont{Heinz}}\ and\ \bibinfo
  {author} {\bibfnamefont{A.}~\bibnamefont{Kuhlman}},\ }%
  \bibfield{journal}{%
  \Doi{10.1103/PhysRevLett.94.132301}{\bibinfo {journal} {Phys. Rev. Lett.}}\
  }%
  \textbf{\bibinfo {volume} {94}},\ \bibinfo {pages} {132301} (\bibinfo {year}
  {2005})%
  \bibAnnoteFile{NoStop}{Heinz:2004ir}%
\bibitem{Goldschmidt:2015kpa}%
  \BibitemOpen
  \bibfield{author}{%
  \bibinfo {author} {\bibfnamefont{A.}~\bibnamefont{Goldschmidt}}, \bibinfo
  {author} {\bibfnamefont{Z.}~\bibnamefont{Qiu}}, \bibinfo {author}
  {\bibfnamefont{C.}~\bibnamefont{Shen}},\ and\ \bibinfo {author}
  {\bibfnamefont{U.}~\bibnamefont{Heinz}},\ }%
  \bibfield{journal}{%
  \Doi{10.1103/PhysRevC.92.044903}{\bibinfo {journal} {Phys. Rev. C}}\ }%
  \textbf{\bibinfo {volume} {92}},\ \bibinfo {pages} {044903} (\bibinfo {year}
  {2015})%
  \bibAnnoteFile{NoStop}{Goldschmidt:2015kpa}%
\bibitem{Rybczynski:2012av}%
  \BibitemOpen
  \bibfield{author}{%
  \bibinfo {author} {\bibfnamefont{M.}~\bibnamefont{Rybczynski}}, \bibinfo
  {author} {\bibfnamefont{W.}~\bibnamefont{Broniowski}},\ and\ \bibinfo
  {author} {\bibfnamefont{G.}~\bibnamefont{Stefanek}},\ }%
  \bibfield{journal}{%
  \Doi{10.1103/PhysRevC.87.044908}{\bibinfo {journal} {Phys. Rev. C}}\ }%
  \textbf{\bibinfo {volume} {87}},\ \bibinfo {pages} {044908} (\bibinfo {year}
  {2013})%
  \bibAnnoteFile{NoStop}{Rybczynski:2012av}%
\bibitem{Haque:2011aa}%
  \BibitemOpen
  \bibfield{author}{%
  \bibinfo {author} {\bibfnamefont{M.~R.}\ \bibnamefont{Haque}}, \bibinfo
  {author} {\bibfnamefont{Z.-W.}\ \bibnamefont{Lin}},\ and\ \bibinfo {author}
  {\bibfnamefont{B.}~\bibnamefont{Mohanty}},\ }%
  \bibfield{journal}{%
  \Doi{10.1103/PhysRevC.85.034905}{\bibinfo {journal} {Phys. Rev. C}}\ }%
  \textbf{\bibinfo {volume} {85}},\ \bibinfo {pages} {034905} (\bibinfo {year}
  {2012})%
  \bibAnnoteFile{NoStop}{Haque:2011aa}%
\bibitem{Chatterjee:2014sea}%
  \BibitemOpen
  \bibfield{author}{%
  \bibinfo {author} {\bibfnamefont{S.}~\bibnamefont{Chatterjee}}\ and\ \bibinfo
  {author} {\bibfnamefont{P.}~\bibnamefont{Tribedy}},\ }%
  \bibfield{journal}{%
  \Doi{10.1103/PhysRevC.92.011902}{\bibinfo {journal} {Phys. Rev. C}}\ }%
  \textbf{\bibinfo {volume} {92}},\ \bibinfo {pages} {011902} (\bibinfo {year}
  {2015})%
  \bibAnnoteFile{NoStop}{Chatterjee:2014sea}%
\bibitem{Adamczyk:2015obl}%
  \BibitemOpen
  \bibfield{author}{%
  \bibinfo {author} {\bibfnamefont{L.}~\bibnamefont{Adamczyk}} \emph{et~al.}
  (\bibinfo {collaboration} {STAR}),\ }%
  \bibfield{journal}{%
  \Doi{10.1103/PhysRevLett.115.222301}{\bibinfo {journal} {Phys. Rev. Lett.}}\
  }%
  \textbf{\bibinfo {volume} {115}},\ \bibinfo {pages} {222301} (\bibinfo {year}
  {2015})%
  \bibAnnoteFile{NoStop}{Adamczyk:2015obl}%
\bibitem{Moreland:2014oya}%
  \BibitemOpen
  \bibfield{author}{%
  \bibinfo {author} {\bibfnamefont{J.~S.}\ \bibnamefont{Moreland}}, \bibinfo
  {author} {\bibfnamefont{J.~E.}\ \bibnamefont{Bernhard}},\ and\ \bibinfo
  {author} {\bibfnamefont{S.~A.}\ \bibnamefont{Bass}},\ }%
  \bibfield{journal}{%
  \Doi{10.1103/PhysRevC.92.011901}{\bibinfo {journal} {Phys. Rev. C}}\ }%
  \textbf{\bibinfo {volume} {92}},\ \bibinfo {pages} {011901} (\bibinfo {year}
  {2015})%
  \bibAnnoteFile{NoStop}{Moreland:2014oya}%
\bibitem{Bozek:2012fw}%
  \BibitemOpen
  \bibfield{author}{%
  \bibinfo {author} {\bibfnamefont{P.}~\bibnamefont{Bo\.zek}}\ and\ \bibinfo
  {author} {\bibfnamefont{W.}~\bibnamefont{Broniowski}},\ }%
  \bibfield{journal}{%
  \bibinfo {journal} {Phys. Rev.}\ }%
  \textbf{\bibinfo {volume} {C85}},\ \bibinfo {pages} {044910} (\bibinfo {year}
  {2012})%
  \bibAnnoteFile{NoStop}{Bozek:2012fw}%
\end{thebibliography}%

\end{document}